\documentclass[twocolumn,amsmath,groupedaddress,resetfootnote]{aastex701}

\makeatletter
\long\def\frontmatter@title@above{%
	\vspace*{-\headsep}\vspace*{\headheight}%
	\par\vspace*{0.25in}%
}
\makeatother

\newcommand{\runinlabel}[1]{\par\medskip\noindent\textit{#1}\hspace{0.4em}\ignorespaces}

\shorttitle{Atmospheric escape fractionates secondary atmospheres}
\shortauthors{Attia \& Lichtenberg}

\def\scititle{
	Atmospheric escape fractionates secondary but not primary atmospheres
}

\begin{document}

\title{\scititle}

\author[orcid=0000-0002-7971-7439,gname=Mara,sname=Attia]{Mara Attia}
\affiliation{Kapteyn Astronomical Institute, University of Groningen, Groningen, Netherlands}
\email[show]{m.attia@rug.nl}

\author[orcid=0000-0002-3286-7683,gname=Tim,sname=Lichtenberg]{Tim Lichtenberg}
\affiliation{Kapteyn Astronomical Institute, University of Groningen, Groningen, Netherlands}
\email{m.attia@rug.nl}

\correspondingauthor{Mara Attia}

\begin{abstract}
A planet's atmosphere, heated by starlight, can flow off as a wind, taking some gases and leaving others in unknown proportions. Yet this selection wrote the noble-gas records of the terrestrial planets, and decides which gases hot exoplanets keep. We solve this fractionated escape problem for arbitrary composition: a wind sorts gases only for secondary, hydrogen-poor atmospheres, and removes primary, hydrogen-rich atmospheres wholesale. The algebraic solution recovers every published formula as special cases. It provides a unified interpretation of atmospheric evolution across the Solar System and exoplanets. Fractionating argon on Mars stripped carbon while sparing krypton, Venus' retained argon constrains the wind intensity that removed its water, Earth's xenon record excludes a neutral wind. The same solution ranks the rocky exoplanets under JWST observation by the gases they can keep.
\end{abstract}

\maketitle
\makeatletter
\global\@firstsectionfalse
\let\footnotetext=\old@foot@note@text
\let\footnotemark=\old@foot@note@mark
\if@two@col\twocolumngrid\@booleantrue\twocolumn@sw\fi
\makeatother

\noindent
A planet whose upper atmosphere absorbs enough of its star's X-ray and extreme-ultraviolet (XUV) radiation does not hold that atmosphere in place: the heated gas flows outward as a hydrodynamic wind~\citep{Watson1981, Kubyshkina2026}. The wind is selective. Light gases escape readily, while heavier ones are dragged along or stay behind. This mass fractionation shaped some of the oldest records in planetary science, the noble-gas and isotope patterns of Venus, Earth, and Mars~\citep{Hunten1987, Sasaki1988, Dauphas2014, Zahnle2023}, and it influences which gases the James Webb Space Telescope (JWST) can find on rocky exoplanets~\citep{Wordsworth2022, Kreidberg2025, Coy2025, Lichtenberg2025}.

Here we show that this selection obeys a sorting condition: a wind fractionates an atmosphere only when the bulk of the atmosphere is dominated by heavy species. The escape thresholds of the gases, the minimal escape flux necessary to drag a gas species out to space, split into two regimes separated by one wide gap. This fractionation gap sits immediately above the escape threshold of the lightest abundant gas in an atmosphere and widens inversely with that gas' abundance (Fig.~\ref{fig:sorting}). Consequently, a hydrogen-rich atmosphere is removed wholesale, while a hydrogen-poor one is stripped gas by gas in a fixed order (Fig.~\ref{fig:ladder}). The fractionation sequence and gap ordering emerge from a general solution of the escape problem, derived here for any number of gases at arbitrary composition and escape strength. Critically, the identity of the escaping gases is an output of the solution rather than a modeling choice. That solution costs no more than evaluating a formula instead of performing costly numerical simulations, and turns into a ranked and falsifiable prediction for the atmospheric composition of the rocky exoplanets in JWST's current emission programs (Fig.~\ref{fig:retention}). Turned toward the Solar System, the general solution demonstrates that early Mars was stripped of most of its carbon, rules out a neutral escape wind as the origin of Earth's xenon record, and bounds the escape flux of early Venus by the argon it keeps (Fig.~\ref{fig:noblegas}). Every published fractionation formula in the literature emerges from the presented solution as a special case, so where two published forms contradict one another our general solution decides between them (Table~\ref{tab:adjudication} and Table~\ref{tab:s_reductions}). We verify the accuracy of our derived escape fractionation solution against an independent transonic multifluid simulation, which reproduces its fractionation factors to a few parts in ten thousand (Fig.~\ref{fig:validation}).

The problem of which gases escape in which order from a planetary atmosphere is at least fifty years old. Rigorous mathematical treatment began with the diffusion-limited flux approach of \citet{Hunten1973} and the crossover mass of \citet{Hunten1987}, the mass above which a gas can no longer be dragged along. Every treatment since has solved a special case for the context and specific planet under study, with the escaping gases specified as a pre-designated input~\citep{Sasaki1988, Chassefiere1996a, Zahnle2023}. The general problem, with all fluxes unknown at once, was written down twice and both times set aside unsolved~\citep{Zahnle1990, Zahnle2023}. Building on these foundations, we here solve this general atmospheric escape fractionation problem through a prescribed total mass flux, the bootstrap condition of multicomponent mass transfer~\citep{Krishna1979, Taylor1993, Chassefiere1996a}. In the solution we derive here, the flux ratios and the set of gases carrying nonzero flux are outputs of the same system. To our knowledge, the non-negativity of escape fluxes has not been handled as a complementarity problem before, in neither the escape nor the multicomponent mass-transfer literature. In both, zero-flux gases are designated in advance rather than found by the solver.

\section*{Closing the multispecies escape problem}

An escaping planetary wind carries $N$ gases with number fluxes $\Phi_j$, base mole fractions $X_j$, and particle masses $m_j$, at temperature $T$ and gravitational acceleration $g_0$, coupled pairwise by binary diffusion parameters $b_{ij}$. The total mass flux $\phi$ comes from any escape-rate prescription~\citep{Watson1981, Erkaev2007, Caldiroli2021}. Eliminating the velocities from the multispecies momentum equations, whose friction terms are the Stefan--Maxwell drag, and applying the shared constant-composition closure turns the system into algebra (materials and methods). In the drift variables $w_j = \Phi_j/X_j$, the solution is the pair $(w, \bar{H})$ satisfying, for every gas $j$,

\begin{equation}
	\begin{aligned}
		\sum_i \frac{X_i\,(w_i - w_j)}{b_{ij}} &= \frac{m_j g_0}{k_\mathrm{B}T} - \frac{1}{\bar{H}} &&\mbox{if } w_j > 0,\\[4pt]
		\sum_i \frac{X_i\,w_i}{b_{ij}} &\le \frac{m_j g_0}{k_\mathrm{B}T} - \frac{1}{\bar{H}} &&\mbox{if } w_j = 0,
	\end{aligned}
	\label{eq:closure}
\end{equation}

\noindent together with the prescribed-flux constraint

\begin{equation}
	\sum_j m_j X_j w_j = \phi, \qquad \Phi_j = X_j w_j \ge 0,
	\label{eq:massflux}
\end{equation}

\noindent where $k_\mathrm{B}$ is the Boltzmann constant. $\bar{H}$, the one density scale height shared by every escaping gas, is itself an unknown of the problem. Each gas either escapes, with the drag from all other gases balancing its weight excess, or stays, when the wind's friction is unable to lift it. Which branch each gas takes is decided by the system rather than by the modeler.

Equations~\eqref{eq:closure} and \eqref{eq:massflux} identify the minimum of a cost function: the wind's friction losses plus the power spent lifting gas out of the gravity well. This function is solved at fixed total particle flux. The cost function's convex shape guarantees one sole minimum (supplementary materials). This structure resolves the questions that kept the system unsolved so far: (i) the solution exists and is unique for any composition and any flux; (ii) the fluxes are continuous and piecewise linear in $\phi$; (iii) each gas has one activation threshold $\phi_k^*$, and the escaping set grows monotonically with flux; and (iv) as $\phi \to 0$ the flux collapses onto the lightest gas, the diffusion-limited configuration of one gas escaping through a static background~\citep{Hunten1973}, whose limiting flux is the lowest of those thresholds.

Every algebraic fractionation formalism published thus far reduces to Eqs.~\eqref{eq:closure} and \eqref{eq:massflux} in its own domain (Table~\ref{tab:adjudication}). The same reductions settle all published algebraic disagreements and close its one remaining open case, with each mechanism derived in the supplementary materials. This general solution costs no more than evaluating an escape-rate formula, under a millisecond even for a 14-component gas atmosphere, against minutes to an hour for one converged steady state of the flow simulation we test against below, let alone a multispecies radiation-hydrodynamic treatment~\citep{Caldiroli2021, Schulik2023MNRAS}.

\begin{table*}[t]
	\centering
	\caption{\textbf{Every printed escape-fractionation formalism is a special case of Eqs.~\eqref{eq:closure} and \eqref{eq:massflux}.} Each recovery is an automated test of the deposited code (materials and methods). $f_2$ and $\mu_2$ are the heavy-to-light mixing and mass ratios.}
	\label{tab:adjudication}

	\small
	\begin{tabular}{p{0.30\textwidth}p{0.28\textwidth}p{0.34\textwidth}}
		\\
		\hline
		Printed formalism & Scope as printed & Under the general solution\\
		\hline
		Diffusion-limited flux~\citep{Hunten1973} & lightest gas through a static background & the light-gas flux at the lowest activation threshold\\
		Crossover mass, Eqs.~(16)--(17) of \citet{Hunten1987} & two gases, one diffusion parameter & recovered to $5\times10^{-15}$; the printed Earth, Mars, and Venus anchors reproduced within their rounding\\
		Continuous two-species fractionation, Eqs.~(14) and (16) of \citet{Zahnle1986} & two gases through the crossover & recovered at crossover; adopted as the near-threshold accuracy bracket\\
		Multicomponent system, Eqs.~(34)--(36) of \citet{Zahnle1986} & many gases & correct only for trace heavies; Eq.~(35) survives by cancellation, Eq.~(36) is off by $(1{+}f_2)/(\mu_2{+}f_2)$, an order of magnitude in O-dominated flows\\
		Two majors plus minors, Eqs.~(35), (36), and (42) of \citet{Zahnle1990} & trace minors over majors escaping (35), at the limiting flux (36), or retained (42) & recovered in the corrected form, including the limiting fluxes\\
		Prescribed-flux binary partition, Eqs.~(1)--(7) of \citet{Chassefiere1996a} & H and non-trace O, dropout by crossover test & recovered to $2\times10^{-14}$\\
		Two majors plus trace minors, Eqs.~(4)--(5) of \citet{Odert2018} & two escaping majors & recovered to $10^{-11}$; their hard zero is the complementarity branch\\
		Binary and isotope systems~\citep{Hu2015, Wordsworth2018, Cherubim2024} & two gases plus isotopes, hard threshold & recovered; the hard threshold discards the measured sub-threshold tail (15\% for D at its threshold)\\
		Ternary isotope system, Eqs.~(4), (8), and (12) of \citet{Gu2023} & H/He/D, designated primary & recovered in the trace limit; their discarded back-reaction term is part of the retained coupling\\
		Mole-fraction system, Eqs.~(9), (19), and (20) of \citet{Zahnle2023} & general $N$ for Eq.~(9); three gases, escaping set chosen by regime & identities at $N = 3$; the all-escaping case, which the authors state can be solved and do not use, is solved here\\
		\hline
	\end{tabular}
\end{table*}

We measured the solution's accuracy against a purpose-built transonic multifluid wind simulation that shares the inherited frame of one temperature, neutral gases, and no chemistry. It tests the critical approximations of the general solution: the subsonic setting of flux ratios and constant co-escaping composition (materials and methods). The comparison is organized by each gas' distance from its own threshold, $z_k = \Lambda_k\,(\phi/\phi_k^* - 1)$. Within this coupled system, the slope $\Lambda_k$ generalizes the product of (i) the restricted Jeans parameter and (ii) the gas' mass excess over the carrier (Eq.~\eqref{eq:s_zdef}). With every gas far from its own threshold, at $z_k \gtrsim 15$, the fractionation factors agree with the simulations to better than $3.5\times10^{-4}$ across two to four gases, mass ratios of 2 to 16, and heavy-gas mass loadings up to about one third. We call this the theory's verified domain (Fig.~\ref{fig:validation}). Below its threshold a gas does not stop escaping at once: the numerical simulations resolve a tail that the general algebraic solution sets to zero by construction. However, this non-resolved tail floors at $\exp(-\Lambda_k)$, and the published two-species continuous form~\citep{Zahnle1986} tracks it within a factor of 1.1 to 2.5, supplying the accuracy bracket and the observability floor wherever a boundary sits near threshold (Fig.~\ref{fig:ladder}B). The hard threshold is kept because the selection structure that follows rests on it (supplementary materials).

\section*{When escape sorts gases}

\begin{figure*}[t]
	\centering
	\includegraphics[width=0.95\textwidth]{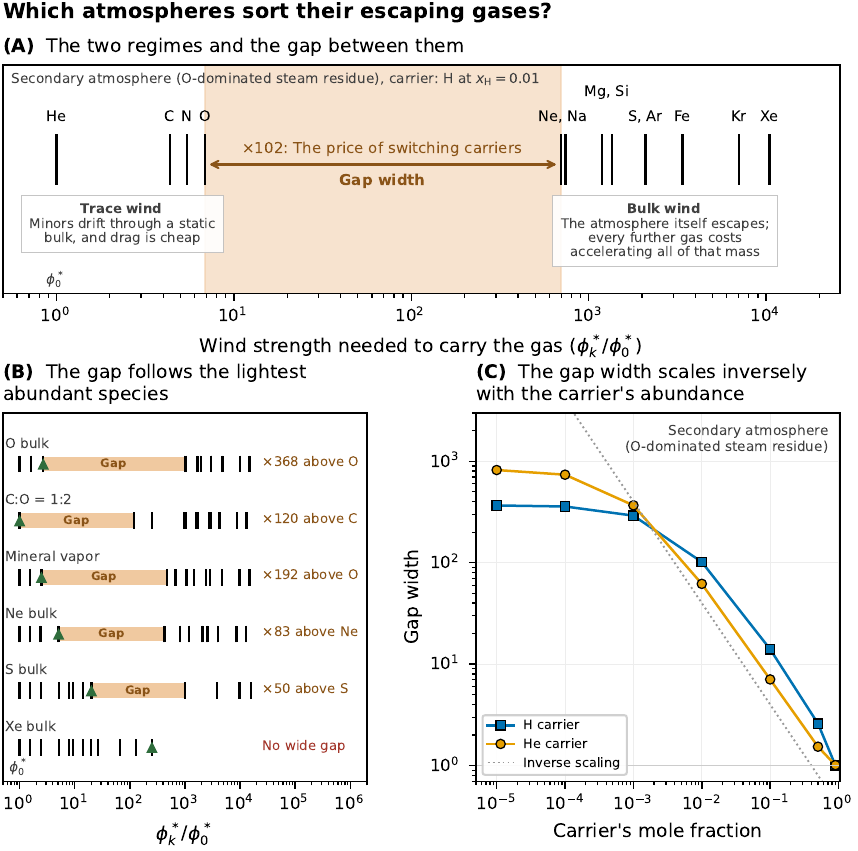}
	\caption{\textbf{Escape sorts gases only when the bulk of the atmosphere is heavy.}
		(\textbf{A})~Wind strengths needed to carry each gas $k$, measured by their entrainment thresholds $\phi_k^*$. Thresholds are normalized relative to the lowest one. The considered secondary atmosphere is dominated by oxygen, as a residue from a lost steam envelope. One wide gap separates a trace-wind regime, in which minor gases drifting through a static bulk, from a bulk-wind regime, in which every further gas is accelerated with the whole atmosphere.
		(\textbf{B})~The gap sits immediately above the lightest abundant gas in every composition whose bulk is heavier than that gas (arrowheads, widths annotated). In the xenon-dominated null case no wide gap forms and the sequence collapses.
		(\textbf{C})~The width of the gap above oxygen against the carrier's mole fraction, for hydrogen and helium carriers, with the inverse-scaling reference (dotted).}
	\label{fig:sorting}
\end{figure*}

The thresholds $\phi_k^*$ carry the fractionation condition. Figure~\ref{fig:sorting}A shows them for what we define here as our fiducial archetype of a secondary, hydrogen-poor atmosphere: a residual steam atmosphere, which is the oxygen-dominated remnant that is left over when stellar irradiation splits water and the ionized hydrogen escapes to space \citep{Luger2015AsBio,Tian2018}. Here, we fix this composition to an oxygen-dominated bulk composition holding one percent of hydrogen. The thresholds order the gases into one sequence with one wide gap, a factor of 102 in flux, immediately above the lightest abundant gas. Below the gap the wind is trace gases drifting through a static bulk, and carrying one more gas is cheap. Above the gap the bulk itself is escaping, and every further gas must be accelerated with all of that mass. The gap is the price of switching carriers, and it follows the lightest abundant gas, with measured widths of 50 to 368 in five example atmospheric compositions (Fig.~\ref{fig:sorting}B). In the xenon-dominated null case, where the bulk is the heaviest gas present, the sequence collapses and no wide gap appears.

The gap's width scales inversely with the carrier's abundance (Fig.~\ref{fig:sorting}C), from a factor of 368 with no hydrogen to 14 at a hydrogen mole fraction of $10^{-1}$. These widths are measured properties of the solution, bounded above by the cost of accelerating the bulk (supplementary materials). The condition for a sorting wind is therefore observable in principle: a heavy bulk, with the light carrier below roughly a percent. A hydrogen-dominated atmosphere cannot sort detectably. Its entire threshold sequence spans a factor of 26, compressed into the coefficient uncertainties. In contrast, the secondary atmosphere spans four decades of flux: escaping winds sort gases only in hydrogen-poor atmospheres. This divide is observed in the exoplanet record: escape thins primary, hydrogen-dominated atmospheres without sorting them, imprinting the radius valley~\citep{Fulton2017, Owen2017, Gupta2019}, while it can compositionally split secondary atmospheres~\citep{LuquePalle2022}.

\section*{A threshold sequence for every atmosphere}

\begin{figure*}[t]
	\centering
	\includegraphics[width=0.95\textwidth]{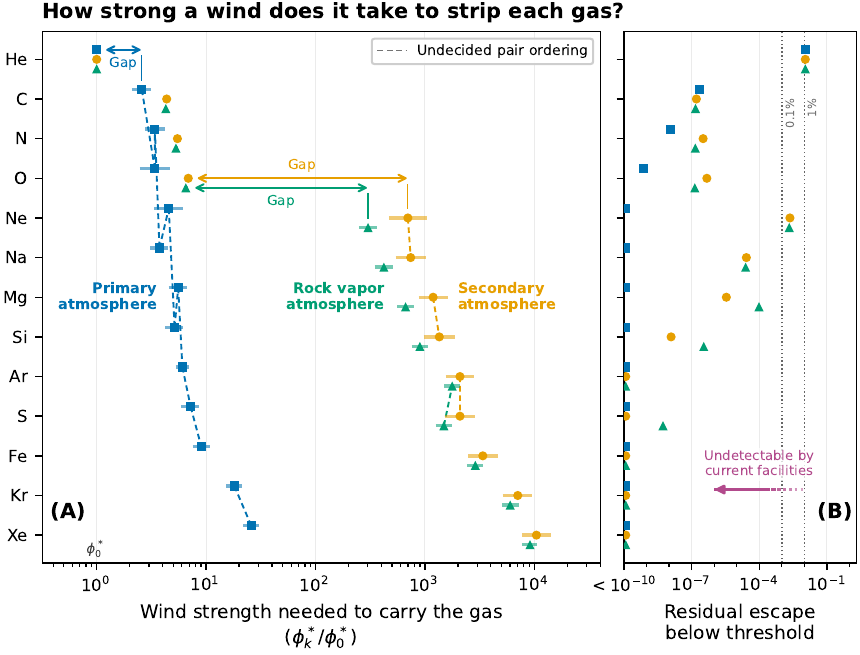}
	\caption{\textbf{The escape threshold sequences of three archetype atmospheres.}
		(\textbf{A})~Wind strengths needed to carry each gas $k$ measured by their thresholds $\phi_k^*$ relative to each atmosphere's lowest threshold. Shown archetypes are a primary atmosphere (blue squares), the secondary atmosphere archetype as in Fig.~\ref{fig:sorting} (orange circles), and a rock vapor (magma-dominated exoplanet, green triangles). Horizontal bands are the correlated-model binary-diffusion coefficient uncertainties. Dashed links connect adjacent pairs that the coefficients leave undecided (text). The two heavy atmospheres sort over four decades, the primary atmosphere sequence is compressed into the coefficient noise.
		(\textbf{B})~Each gas' residual escape below its entrainment threshold. This quantifies the error of the derived solution: the amount of gas lost if the received stellar flux is below the entrainment threshold, which decreases exponentially with $\Lambda_k$ (text and Eq.~\eqref{eq:s_zdef}). A threshold is an observational boundary only if its pair is decided and its residual sub-threshold escape is undetectable (pink arrow).
        }
	\label{fig:ladder}
\end{figure*}

Figure~\ref{fig:ladder} draws the sequence for a primary atmosphere, the fiducial secondary atmosphere, and a rock vapor atmosphere. In the two heavy archetypes the escape order is $\mathrm{H} < \mathrm{He} < \mathrm{C} < \mathrm{N} < \mathrm{O} < \{\mathrm{Ne}, \mathrm{Na}\} < \{\mathrm{Mg}, \mathrm{Si}\} < \{\mathrm{S}, \mathrm{Ar}\} < \mathrm{Fe} < \mathrm{Kr} < \mathrm{Xe}$, with the braced pairs undecided. Rock-forming elements interleave along one sequence rather than separating into a block of their own. The ordering is set by the masses and the diffusion coefficients alone, and moves with neither the well depth, the base radius, the temperature, nor the escape-rate prescription. Its uncertainties are set by the available coefficient library, whose entries are mostly scaled from the same few empirically derived numbers, so that ratios are better determined than values. Under the correlated error model (supplementary materials), 3 of 78 gas pairs are undecided in the secondary atmosphere (Mg/Si, Ne/Na, and S/Ar, with 9 of 12 adjacent pairs separated), 1 of 78 in the rock vapor, and 27 of 78 in the primary atmosphere, whose compressed sequence the coefficients cannot order. An independent-per-row model would leave 13 of 78 undecided in the secondary atmosphere.

The sequence therefore progresses monotonically from light to heavy species, which excludes 75 of the 78 ordered retained--lost pairs in the secondary atmosphere archetype, 77 of 78 in the rock vapor, and 51 of 78 in the primary atmosphere. Not every threshold is a usable boundary, though. The sulfur and argon thresholds coincide to 0.05\% on the secondary atmosphere composition, and act as one boundary (Fig.~\ref{fig:ladder}). Furthermore, a gas below its threshold is not perfectly retained: its escape floors at the $\exp(-\Lambda_k)$ measured above, so a threshold separates observable outcomes only where the floor is beneath the resolvable mixing ratio. Figure~\ref{fig:ladder} prints each floor at a stated 5000~K, inside the 3500 to $10^4$~K span of published thermosphere models~\citep{Kite2020, Ito2021, Cherubim2026}.

\section*{Ranked predictions for observed rocky planets}

\begin{figure*}[t]
	\centering
	\includegraphics[width=\textwidth]{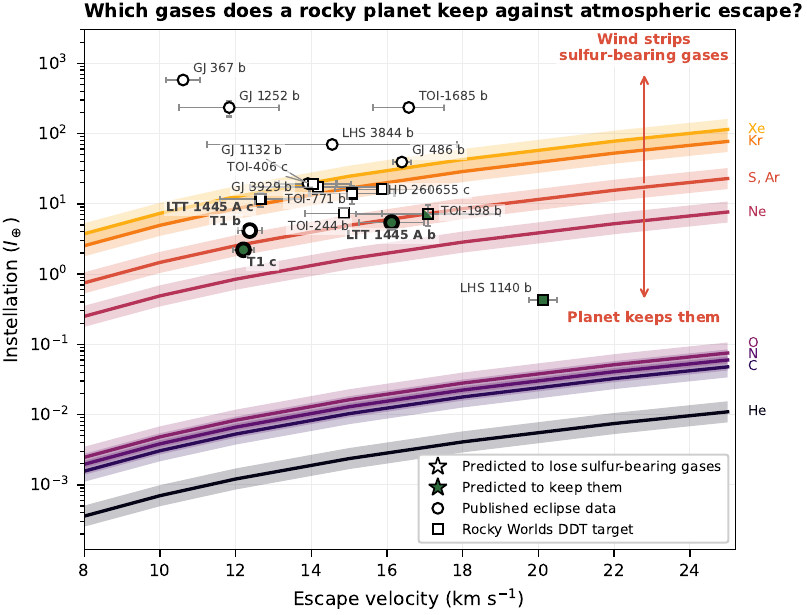}
	\caption{\textbf{Prediction for the atmospheric gas retention of seventeen low-mass exoplanets receiving observational focus with JWST.}
		Each curve is the boundary above which a planet's wind strips that gas from the secondary atmosphere, in instellation against escape velocity. Shaded bands span the adopted range of stellar XUV output. Circles are all published eclipse data of low-mass exoplanets~\citep{Coy2025}. Squares show the upcoming targets of the Rocky Worlds DDT survey. Filled symbols are predicted to keep their sulfur-bearing gases, open symbols to lose them. Error bars are 1$\sigma$ (Table~\ref{tab:s_planets}). TRAPPIST-1~b and c and LTT~1445~A~c and b, each pair crosses the boundary and they are labeled in bold. Escape-rate prescriptions carry an uncertainty of a factor of 12 in flux, which can shift the whole boundary family together.}
	\label{fig:retention}
\end{figure*}

An escape-rate prescription converts each threshold into a boundary in instellation against escape velocity, above which a planet's wind strips that gas (Fig.~\ref{fig:retention}). The boundary family derived here reduces to the energy-limited prediction for the proposed ``cosmic shoreline''~\citep{Zahnle2017}, which ties the critical XUV instellation to escape velocity and bulk density as $v_\mathrm{esc}^3\sqrt{\rho}$, and adds what a single shoreline cannot carry: species resolution. That scaling needs no fitting, and the two routes share no physics beyond the escape-rate prescription: \citet{Zahnle2017} fixed the removed mass fraction, our solution fixes an entrainment threshold proportional to surface gravity, and both give a critical flux proportional to $\rho r_0$ for different reasons (supplementary materials). The prescription factors multiply every boundary together, a bracket of a factor of about 12 in absolute flux, while the spacing between the boundaries is prescription-free (supplementary materials).

Seventeen rocky planets in JWST's current thermal-emission programs, the nine of the published emission sample~\citep{Coy2025} and the nine Rocky Worlds DDT targets~\citep{Redfield2024}\footnote{\url{https://rockyworlds.stsci.edu/}}, one planet (LTT~1445~A~b) belonging to both, land on both sides of the sulfur boundary (Table~\ref{tab:s_planets}). Sulfur is the decisive species for this exoplanet sample, because it is the heaviest element a secondary atmosphere holds in abundance. Its rock-forming neighbors in the sequence fall out by condensation or ionization (supplementary text). Sulfur-bearing gases are moreover observable with current facilities: SO$_2$ has already been shown to explain JWST spectra of WASP-39~b~\citep{Tsai2023} and L~98-59~d~\citep{Nicholls2026}. To quantify the sequence, we establish the escape retention index, a planet's flux margin against its escape boundary. This metric ranks the sample from GJ~367~b at 244, through TRAPPIST-1~b at 1.5, down to LHS~1140~b at 0.035, with TOI-198~b within its instellation uncertainty of the boundary, counted on neither side (Table~\ref{tab:s_planets}).

Escape is not the only writer of the outcome: interior outgassing can resupply what a wind removed, and because redox histories are expected to differ from planet to planet \citep{Wordsworth2022, Lichtenberg2025}, resupply does not shift the sample coherently. A single planet holding a gas it should have lost points to resupply, and only the pattern across the sample tests the escape physics. Planets sharing one star are decisive test cases, having experienced the same irradiation: TRAPPIST-1~b and c cross the sulfur boundary, as do LTT~1445~A~c and b, and the general solution strips each inner planet's sulfur and spares the outer. The reverse outcome, the inner planet sulfur-bearing while the outer is bare, would need interior resupply on one planet and a non-escape loss on the other at once. The prediction also assumes the wind stays a collisional fluid, checked against the published criterion~\citep{Volkov2011, Johnson2013a} in the supplementary materials (Table~\ref{tab:s_regime}).

\section*{The noble-gas record of the Solar System}

The noble-gas records of Venus, Earth, and Mars were written by heavy winds, through the same ranked sequence as Fig.~\ref{fig:retention} for exoplanets, in a regime where previous approaches required a pre-designated composition. Our solution treats them with all gases coupled. The driver of these winds is independently constrained: sodium and potassium lost from the lunar regolith~\citep{Saxena2019} and the neon and argon of Venus and Earth both favor a slowly rotating young Sun~\citep{Lammer2020}. 

\begin{figure*}[t]
	\centering
	\includegraphics[width=0.90\textwidth]{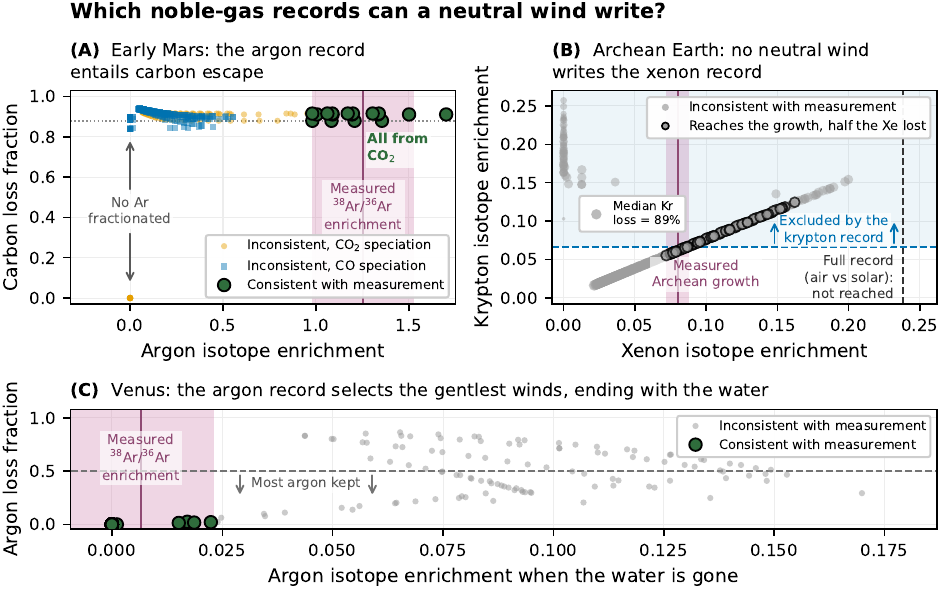}
	\caption{\textbf{What the noble-gas records of Mars, Earth, and Venus entail under the general solution.} Green marks the histories consistent with a measured record, gray those that are not.
		(\textbf{A})~Early Mars: the converged histories of the scan (materials and methods), placed by argon isotope enrichment (the fractional change $E = R/R_0 - 1$ of the heavy-to-light isotopes), for both carbon speciations. The purple band is the meteorite measurement with its 1$\sigma$ width~\citep{Willett2022}. Green points reproduce it, at or beyond the band's lower edge with at least half the argon lost, and all remove at least 88\% of the carbon and none of the krypton. Every one has CO$_2$ as its carbon speciation: no CO history reaches the band. 
		(\textbf{B})~Archean Earth: the same procedure on five Archean compositions, placed by xenon against krypton isotope enrichment ($^{136}$Xe/$^{130}$Xe and $^{86}$Kr/$^{82}$Kr pairs), with marker size scaling as the krypton removed. Nothing here is consistent with the record, which requires the band's xenon growth alongside krypton identical to modern~\citep{Avice2018}. Circled points reach that growth, at or beyond the band's lower edge, with at least half the xenon lost. Every one removes at least 77\% of the krypton and fractionates the rest by at least 13 permil per amu, where the same samples exclude 16 (blue dashed line, shaded above). The remaining points fractionate xenon without losing half of it, and none reaches the full air-to-solar fractionation (black dashed line; \citealt{Dauphas2003}).
		(\textbf{C})~Venus: the same procedure on four water-derived compositions~\citep{Zahnle2023}, each stopped when its hydrogen and oxygen are gone and the wind loses its source. Green points keep argon inside the measured band~\citep{Avice2022} with at most half of it lost (dashed line): they entrain at most 3\% of the argon and none of the CO$_2$. The rest miss the record at higher flux (supplementary materials).}
	\label{fig:noblegas}
\end{figure*}

The $^{38}$Ar/$^{36}$Ar ratio from the meteorite NWA~7034 implies atmospheric escape at 4.4~Ga~\citep{Willett2022, Willett2023}. A hydrogen wind carrying the products of water and carbon dioxide has been suggested qualitatively, but left unsolved~\citep{Zahnle2023}. Solving the system with our general solution (Fig.~\ref{fig:noblegas}A) places carbon's entrainment threshold below krypton's in every composition, carbon speciation, and coefficient-uncertainty model we tested ($P \ge 0.994$). Every converged history in our simulations that (i) reaches the measured argon enrichment within its 1$\sigma$ width, while (ii) losing at least half the argon, removes at least 88\% of the atmosphere's carbon and none of its krypton. The Martian escape wind that set the argon record was therefore stripping carbon, consistent with the weakly fractionated krypton of the young and modern Martian records~\citep{Swindle1986, Conrad2016} (supplementary text).

Earth's atmospheric xenon is isotopically heavy by about 4\% per amu and Mars' by about 2.5\%. Their krypton signature carries no fractionation attributable to escape~\citep{Zahnle2019, Conrad2016}, and Archean samples show Earth's fractionation still growing until about 2.1~Ga~\citep{Avice2018,Parai2018}, long after any primordial wind~\citep{Sasaki1988} or accreted mixture~\citep{Dauphas2003} reset the ratio. No neutral wind solved here produces that pair (Fig.~\ref{fig:noblegas}B). Krypton's escape threshold sits below xenon's ($P = 0.996$ in the secondary atmosphere, 1.000 in the rock vapor, and 0.935 in the primary atmosphere, marginally below our bar of 0.95) whatever the well depth, temperature, or prescription. Therefore, a wind that reaches xenon already carries krypton, as qualitatively suggested by \citet{Zahnle2019}. Below its threshold xenon leaks nothing, its floor sits under $10^{-60}$ at every Archean frame (supplementary materials). Above the threshold no simulated history reaches the full measured fractionation. Every history that reaches the measured Archean growth removes at least 77\% of the krypton and fractionates the rest by at least 13 permil per amu. No simulated history reproduces Archean and modern xenon together with the modern krypton record~\citep{Avice2018}. Earth's xenon record therefore cannot be explained by a neutral wind and instead points to the ion-coupled escape channel~\citep{Zahnle2019, Catling2020}, in which xenon alone among the noble gases is ionized and dragged out by the hydrogen wind along field lines open to space. The theory now clearly rules out the neutral background scenario.

Venus at first appears as a counter-example to fractionated escape: it lost a heavy, water-derived atmosphere~\citep{Hamano2013Natur, Hamano2025, Zahnle2023} yet keeps abundant argon whose isotopes are consistent with solar composition within the measured uncertainties~\citep{Avice2022}. Venus' escape wind seemingly did not sort. In the general solution, fractionation is a near-threshold signature: a gas far below its threshold is spared and its escape floored at the exponentially small $\exp(-\Lambda_k)$, while one far above is carried wholesale. A heavy bulk is necessary for a wind to sort, but not sufficient. Venus' abundant, unfractionated argon therefore marks a wind that removed almost none of it. In four water-derived compositions, every history that strips hydrogen and oxygen while keeping argon inside the measured band entrains at most 3\% of the argon and none of the CO$_2$. Every simulated history at higher flux fractionates or removes the argon. Continued escape after the water is gone strips argon, because argon is lighter than the CO$_2$ left behind (supplementary materials).

These records each fix one property of their planetary escape winds: Mars' composition, Earth's channel, and Venus' flux range. One coordinate carries the resulting constraints we set: the wind strength required, in units of that planet's own argon entrainment threshold. The Martian histories crowd that threshold, at 0.70 to 1.10 of it, the Venusian ones stay below 0.58, and Earth's xenon is written at no value of it.

\section*{One algebra for both records}

Which gases escape and which stay is an output of the general solution rather than an input. Planetary evolution models can fractionate escape at every timestep, for arbitrary volatile compositions, where before they inherited binary formulas and a designated carrier~\citep{Gu2023, Cherubim2024}. Designating the wrong carrier costs a measured factor of about 2000 on a published planet configuration (supplementary text). The coupled interior--atmosphere models built to interpret the coming decade of rocky-planet observations produce atmospheres of growing compositional diversity~\citep{Lichtenberg2021, Nicholls2024, Lichtenberg2026PSJ}, and only an algebraic solution can fractionate them at evolutionary timescales, at under a millisecond per evaluation. The cosmic shoreline, the proposed single boundary between airless and atmosphere-bearing planets~\citep{Zahnle2017, Ji2025}, splits into a mass-ordered sequence of per-gas boundaries, and the question changes from whether a planet holds an atmosphere to which one it holds. One algebra ties the retrospective record to the prospective one. With the young Sun's activity empirically constrained~\citep{Saxena2019, Lammer2020} and the general solution presented here, the escape fluxes reconstructed from the noble gases of Venus, Earth and Mars now form chronologies. The seventeen rocky planets of Fig.~\ref{fig:retention} are already on JWST's observing schedule and test the predicted sequence gas by gas. Ion--ion drag coefficients are known and far larger than binary diffusion coefficients~\citep{Zahnle2019}. Therefore a future extended ionized version of the general solution would further enable testing against Earth's nine xenon isotopes.


\section*{Acknowledgments}
\runinlabel{Funding:}
M.A. is supported by the Swiss National Science Foundation through the Postdoc.Mobility fellowship, grant number 230229. This research was funded by the European Union (ERC, MagmaWorlds, 101219807). Views and opinions expressed are however those of the author(s) only and do not necessarily reflect those of the European Union or the European Research Council. Neither the European Union nor the granting authority can be held responsible for them. T.L. was further supported by the Branco Weiss Foundation, the Alfred P. Sloan Foundation (AEThER, G-2025-25284), NASA’s Nexus for Exoplanet System Science research coordination network (Alien Earths, 80NSSC21K0593), and the NWO NWA-ORC PRELIFE Consortium (NWA.1630.23.013).
\runinlabel{Author contributions:}
M.A. derived the closure, wrote the reference implementation, the validation integrator, and the analysis code, ran the validation and the target-sample analyses, and wrote the manuscript. T.L. reviewed the derivation and its provenance, identified the target sample, and contributed to the manuscript.
\runinlabel{Competing interests:}
There are no competing interests to declare.
\runinlabel{Data, code and materials availability:}
The reference implementation of the closure with its automated tests, the validation simulation with every stored run, the analysis code with every stored result, and the diffusion-coefficient library are deposited at Zenodo (\url{https://doi.org/10.5281/zenodo.22181501}) under an Apache-2.0 license. All other data are available in the main text or the supplementary materials. No physical materials were generated in this work.


\section*{Supplementary materials}
Materials and Methods\\
Supplementary Text\\
Fig.~S1\\
Tables S1 to S5\\
Caption for Data S1


\bibliography{references}
\bibliographystyle{aasjournal}



\renewcommand{\thefigure}{S\arabic{figure}}
\renewcommand{\thetable}{S\arabic{table}}
\renewcommand{\theequation}{S\arabic{equation}}
\renewcommand{\thepage}{S\arabic{page}}
\setcounter{figure}{0}
\setcounter{table}{0}
\setcounter{equation}{0}
\setcounter{page}{1}


\clearpage
\onecolumngrid
\begin{center}
	{\LARGE\bfseries Supplementary Materials\par}
\end{center}
\vspace{18pt}
\twocolumngrid


\section*{Materials and Methods}

\subsection*{The multispecies wind system and its reduction to algebra}

The starting point is the multispecies subsonic wind system of \citet{Zahnle1990}, their Eqs.~(13) to (15): per-species continuity, $n_j u_j r^2 = \Phi_j r_0^2$, and the momentum balance in which each gas' partial-pressure gradient balances its weight and the friction exerted by every other gas,

\begin{equation}
	\frac{k_\mathrm{B}T}{n_j}\frac{\mathrm{d}n_j}{\mathrm{d}r} = -\frac{G M m_j}{r^2} + \sum_i \left(u_i - u_j\right) n_i \frac{k_\mathrm{B}T}{b_{ij}},
	\label{eq:s_momentum}
\end{equation}

\noindent where $n_j$ and $u_j$ are number densities and velocities, $G$ is the gravitational constant, $M$ the planet's mass, and $b_{ij} = n D_{ij}$ the binary diffusion parameter of the pair, the product of the total number density $n$ and the binary diffusion coefficient $D_{ij}$, which is independent of pressure and is the quantity the laboratory compilations tabulate. The friction term is the Stefan--Maxwell drag of multicomponent diffusion theory: the tabulated $b_{ij}$ enter with no composition weighting, an identification that carries the corrections below. All quantities are evaluated at the base level $r_0$, the level where the atmosphere absorbs the escape-driving XUV flux, with the composition atomized there. The total mass flux $\phi$ (per unit area at $r_0$) is an input, supplied by whatever escape-rate prescription is chosen, and the outputs are the per-species number fluxes $\Phi_j \ge 0$ with $\sum_j m_j \Phi_j = \phi$ identically.

Eliminating the velocities with continuity and differencing any two species removes the reference-species asymmetry of the printed forms. The closure assumption, shared by the entire algebraic family~\citep{Hunten1987, Zahnle1986, Zahnle1990, Odert2018, Cherubim2024}, is that composition ratios of co-escaping species are constant with altitude, which \citet{Zahnle1990} justify in the limit that a constituent escapes at a nonnegligible rate. Under it, all co-escaping species share one logarithmic density gradient $-1/\bar{H}$ at the base, every term in Eq.~\eqref{eq:s_momentum} scales as $r^{-2}$, so a solution at $r_0$ holds at all radii, and the system becomes algebraic. In the drift variables $w_j = \Phi_j / X_j$, equal at $r_0$ to $n\,u_j$, the drift velocity of gas $j$ weighted by the total number density,

\begin{equation}
	\sum_i \frac{X_i \left(w_i - w_j\right)}{b_{ij}} = \frac{m_j g_0}{k_\mathrm{B}T} - \frac{1}{\bar{H}} \qquad (j = 1, \ldots, N).
	\label{eq:s_system}
\end{equation}

\noindent Fractionation factors are ratios of drift variables, $\chi_j = w_j / w_\mathrm{ref}$, with the reference gas the lightest one unless stated otherwise: $\chi_j$ is the rate at which the wind removes gas $j$ relative to its abundance, normalized to the reference gas, so $\chi_j = 1$ means gas $j$ escapes in proportion to its abundance and is not fractionated, and $\chi_j = 0$ means it stays behind.

\subsection*{The fully entrained solution}

When every gas escapes, multiplying Eq.~\eqref{eq:s_system} by $X_j$ and summing cancels the drag pairwise and pins the shared scale height at the mean-molecular-mass value,

\begin{equation}
	\frac{1}{\bar{H}} = \frac{\bar{m}\, g_0}{k_\mathrm{B}T}, \qquad \bar{m} = \sum_j X_j m_j,
	\label{eq:s_gradient}
\end{equation}

\noindent so Eq.~\eqref{eq:s_system} becomes a reference-free force balance: the net drag per particle on gas $j$ equals its weight excess over the mean, $(m_j - \bar{m})\,g_0$. Gases lighter than the mean are dragged back by the rest, heavier ones are dragged forward. There is no designated primary anywhere. We build here on two precursor concepts, the binary mean-mass scale height (Eq.~(6) of \citealt{Hunten1987}) and the general-$N$ mole-fraction restatement of the system with the mean-mass driving term (Eq.~(9) of \citealt{Zahnle2023}). Our addition here is the identification of $1/\bar{H}$ as a multiplier that survives when not all mixing-ratio gradients vanish, which is what makes the dropout regime below tractable.

Writing $(\mathsf{L}w)_j = \sum_{i \ne j} (X_i X_j / b_{ij})(w_i - w_j)$, the matrix $\mathsf{L}$ is a symmetric weighted graph Laplacian (gases as nodes, the positive $X_i X_j / b_{ij}$ as edge weights): negative semidefinite, with null space the constant vector. The compatibility condition $\sum_j X_j (m_j - \bar{m}) = 0$ holds identically, so the system determines $w$ up to a constant and the prescribed flux fixes the constant. Existence and uniqueness follow for any composition with positive mole fractions and finite $b_{ij}$, by one linear solve of size $N{+}1$. The solution splits as $w_j = \lambda + v_j$ with $v$ flux-independent: each gas' flux is its mole-fraction share of a common advective flux plus a differential-diffusion correction, the $N$-species generalization of the two-term structure of \citet{Wordsworth2018}. With a single diffusion parameter $b$, the system collapses to a closed form linear in mass, with a single crossover mass $m_\mathrm{c} = \bar{m} + k_\mathrm{B}T\,\Phi_\mathrm{tot}/(b g_0)$, where $\Phi_\mathrm{tot} = \sum_j X_j w_j$ is the total number flux, measured from the mean mass: the two-straight-lines picture of \citet{Hunten1987} generalized to $N$ species, reducing to their Eq.~(16) in the trace-heavy limit. For general $b_{ij}$ there is no single crossover mass, confirming the caveat of \citet{Zahnle1990} that the concept is neither uniquely defined nor especially useful in multicomponent escape: the dropout order is set by drag-adjusted masses, not masses alone.

\subsection*{Species dropout as a complementarity problem}

A gas with $w_k = 0$ sits in a drag-modified hydrostatic distribution, the wind's friction partially supporting its weight. Retention is self-consistent only if its density gradient is at least as steep as the escaping gases' shared $-1/\bar{H}$. Otherwise its mixing ratio would grow with altitude and it would be lofted into the wind. The retention criterion is

\begin{equation}
	R_k = \sum_{i \in \mathcal{A}} \frac{X_i w_i}{b_{ik}} - \left(\frac{m_k g_0}{k_\mathrm{B}T} - \frac{1}{\bar{H}}\right) \le 0 \qquad (k \notin \mathcal{A}),
	\label{eq:s_retention}
\end{equation}

\noindent with $\mathcal{A}$ the escaping set. For $j \in \mathcal{A}$, Eq.~\eqref{eq:s_system} holds with retained species contributing pure drag and $\bar{H}$ no longer pinned at the mean-molecular value, part of the wind's momentum being spent on the static species. Together with $\sum_j m_j X_j w_j = \phi$ this is the closure stated in the main text (Eqs.~\eqref{eq:closure} and \eqref{eq:massflux} there), and it is the Karush--Kuhn--Tucker system of the convex quadratic program

\begin{equation}
	\begin{aligned}
		\min_{w \ge 0}\; &\frac{1}{2}\sum_{i<j} \frac{X_i X_j}{b_{ij}} \left(w_i - w_j\right)^2 + \frac{g_0}{k_\mathrm{B}T}\sum_j m_j X_j w_j\\[2pt]
		&\mbox{subject to} \quad \sum_j X_j w_j = \nu,
	\end{aligned}
	\label{eq:s_qp}
\end{equation}

\noindent friction dissipation plus gravitational power minimized at fixed total number flux, with $1/\bar{H}$ the multiplier of the constraint and $\nu$ tuned so the mass flux equals $\phi$. On the constraint slice (the compact simplex $w \ge 0$, $\sum_j X_j w_j = \nu$) the objective is strictly convex, so the minimizer, the escaping set, and the multiplier are unique. Four consequences follow, each of which we verified numerically:
\begin{enumerate}
\item existence and uniqueness for any $\phi > 0$, verified against exhaustive enumeration of all $2^N - 1$ candidate escaping sets;
\item continuity and piecewise linearity of $w(\phi)$, verified by two-sided evaluation at 113 activation thresholds located by bisection;
\item componentwise monotonicity, $\partial w_j / \partial \phi \ge 0$, via the grounded-Laplacian M-matrix structure on each escaping set, so that each gas has a unique activation threshold $\phi_k^*$ and the escaping set grows monotonically with flux;
\item collapse of the entire flux onto the lightest gas as $\phi \to 0^+$, verified down to $10^{-12}$ of the lowest threshold.
\end{enumerate}

The implementation is an active-set iteration: solve on the current set, remove species with negative drift, re-solve, then re-admit the worst retained violator of Eq.~\eqref{eq:s_retention}, one per pass. The tolerances are part of the algorithm: a species is removed only when its drift falls below a tolerance scaled to $\phi/\min_j m_j$, re-admitted only when its retention residual exceeds one scaled to $m_k g_0/k_\mathrm{B}T$, and within-tolerance negative drifts are set to zero so the returned fluxes satisfy $w \ge 0$. Tested at thresholds located to machine precision in 400 random systems with $N = 2$ to 6, the iteration terminated in every case. Uniqueness of the escaping set holds for every $\phi$ off the finite set of thresholds. The cost for $N \lesssim 10$ is a handful of linear solves of size at most $N{+}1$: measured at about 0.4~ms per 14-species solve and 0.27~s per complete 13-threshold sequence (one workstation core, plain Python reference implementation, order-of-magnitude figures), the sub-millisecond cost the main text quotes.

\subsection*{The near-threshold tail}

The hard zero below threshold inherits a known approximation of the algebraic family: continuous treatments~\citep{Zahnle1986, Zahnle1990} keep a small nonzero escape efficiency near crossover, of relative size $1/(1 + \Lambda_k)$ at the threshold. The organizing coordinates are

\begin{equation}
	z_k = \Lambda_k \left( \frac{\phi}{\phi_k^*} - 1 \right), \qquad \Lambda_k = r_0 \left. \frac{\mathrm{d}R_k}{\mathrm{d}\ln\phi} \right|_{\phi_k^*},
	\label{eq:s_zdef}
\end{equation}

\noindent the wind's distance from gas $k$'s threshold and the slope of the closure's own retention residual there. In a two-gas atmosphere $\Lambda_k = \lambda_0(\mu_k - 1)$, with $\lambda_0 = G M m_1/(k_\mathrm{B}T r_0)$ the restricted Jeans parameter and $\mu_k = m_k/m_1$, so $\Lambda_k$ generalizes the product of well depth and mass excess to the coupled system. Rewritten in these variables, Eq.~(14) of \citet{Zahnle1986} becomes

\begin{equation}
	\chi(z, \Lambda) = \frac{z/\Lambda}{1 + z/\Lambda - e^{-z}},
	\label{eq:s_tail}
\end{equation}

\noindent with $\chi$ the entrained fraction relative to the carrier defined above. It passes through $1/(1+\Lambda)$ at $z = 0$ and floors at $e^{-\Lambda}$ at $z = -\Lambda$, the zero-flux end of the coordinate. Against the converged sub-threshold validation measurements it is low by factors of 1.13 to 1.96 in the three-species mixtures and 1.09 to 2.45 in the four-gas control, with a bias that is a mild function of $\Lambda$ alone. The measured tail is not a function of $(z, \Lambda)$ alone: at fixed $z$ the residual shifts by up to a factor of 2.19 between heavy-gas loadings, so no closed-form tail in these coordinates can be exact, and $z$ decides inclusion of the verified domain but never interpolates an error inside the unverified band.

The hard threshold is kept in the closure for structural reasons. The tail is a property of variation with altitude, and the closure is an algebra problem at one altitude: a species near its threshold is the one that does not share the common density falloff, and restoring the composition-gradient term that carries this means integrating with height, forfeiting the requirement that fractionation cost no more than an escape-rate evaluation. And the hard zero is one branch of the complementarity conditions: uniqueness, continuity, piecewise linearity, and one threshold per species all follow from it, whereas under any smooth tail every species escapes at least a little at any total flux, and a threshold becomes an observer's convention. Where a boundary sits near threshold, Eq.~\eqref{eq:s_tail} is quoted alongside as the accuracy bracket and the observability floor. The radial drift of a retained species' mixing ratio is bounded within the framework by $\exp[r_0 R_k (1 - r_0/r)]$, vanishing at the activation threshold. Sasaki and Nakazawa reached the same structure in the binary from the opposite direction and named the near-threshold exception, a species within about $0.1\,m_\mathrm{H}$ of the crossover mass that may ooze~\citep{Sasaki1988}.

\subsection*{Exact reductions and the corrections}

Table~\ref{tab:s_reductions} lists the printed formalisms recovered from the closure, the configuration of each recovery, and the measured residuals. Every row is one of the deposited automated tests, each of which recomputes its target and fails on any residual above its stated tolerance. The corrections quoted in the main text are established as follows. The multicomponent system of \citet{Zahnle1986} (their Eqs.~(34) to (36)) builds the drag with a composition-weighted collision rate whose per-particle force coefficient equals the Stefan--Maxwell value $k_\mathrm{B}T/b_{ij}$ only when species $j$ is trace or the masses are equal. Solving their own two-major subsystem, the spurious weights enter the two momentum equations as factors $\alpha = (1 + f_2)/(\mu_2 + f_2)$ and $\mu_2 \alpha$. In the difference that eliminates the common gradient they combine to the correct coefficient, so their Eq.~(35) survives by cancellation, but the common gradient itself comes out wrong by $\alpha$ and their minor-species Eq.~(36) inherits it on the buoyancy term, roughly an order of magnitude for oxygen-dominated flows ($\mu_2 = 16$, $f_2 \sim 0.5$ gives $\alpha \approx 0.09$). This substantiates, with a mechanism, the bare statement of \citet{Zahnle1990} that those equations are algebraically incorrect, and it resolves the printed disagreement between Eq.~(36) of \citet{Zahnle1986} and Eq.~(5) of \citet{Odert2018}: the Odert forms are the correct constant-composition reductions, recovered from the closure wherever the minor is entrained, with the closure returning zero where the printed form goes negative.

\begin{table*}[t]
	\centering
	\caption{\textbf{The reductions: printed formalisms recovered from the closure.} Each row is verified by the deposited automated tests. Residuals are maxima over the tested configurations. Trace-species formulas are compared in the strict trace limit, being truncations at trace abundance.}
	\label{tab:s_reductions}

	\small
	\begin{tabular}{p{0.42\textwidth}p{0.28\textwidth}p{0.22\textwidth}}
		\\
		\hline
		Recovered target & Configuration & Residual\\
		\hline
		Crossover mass and binary partition, Eqs.~(16)--(17) of \citet{Hunten1987} & two gases, both escaping & $< 5\times10^{-15}$\\
		Worked anchors of \citet{Hunten1987} (Earth $m_\mathrm{c} = 140$ at $8.19\times10^{13}$; Mars 130 at $2.89\times10^{13}$; Venus $m_\mathrm{c} = 1.356$) & routed through the solver's own threshold & within the printed rounding (2\%)\\
		Binary partition and critical flux, Eqs.~(7)--(9) of \citet{Cherubim2024}; Eqs.~(9)--(12) of \citet{Wordsworth2018} & two gases, both branches, continuous at $\phi_\mathrm{c}$ & $< 5\times10^{-15}$\\
		Prescribed-flux partition, Eq.~(6) of \citet{Chassefiere1996a} & H with non-trace O, 800 random draws & $2.1\times10^{-14}$\\
		Two majors plus trace minor, Eq.~(5) of \citet{Odert2018} = Eq.~(35) of \citet{Zahnle1990}; Eq.~(36) of \citet{Zahnle1990} at the limiting flux & two escaping majors, trace minor & $< 10^{-11}$ (strict trace limit)\\
		Two-background thresholds, Eq.~(42) of \citet{Zahnle1990} & one escaping gas, two retained heavies, including the entrainment order of the two & $< 10^{-11}$\\
		Ternary isotope system, Eqs.~(4), (8), and (12) of \citet{Gu2023}; Eq.~(11) of \citet{Cherubim2024} & H/He/D, He escaping or retained, continuous at the He crossover & $< 10^{-11}$ (strict trace limit)\\
		Crossover fractionation value, Eq.~(16) of \citet{Zahnle1986} & two gases at threshold & recovered; adopted as the tail bracket (Eq.~\eqref{eq:s_tail})\\
		Multibackground threshold, Eq.~(19) of \citet{Zahnle2023} & H escaping over retained O and CO$_2$, all non-trace & $3.9\times10^{-16}$\\
		Difference relation, Eq.~(20) of \citet{Zahnle2023} & H and O escaping over retained CO$_2$, 36 points & $1.0\times10^{-15}$\\
		Published crossover flux of \citet{Kite2020} & binary activation threshold, independent unit conversion & $10^{-12}$\\
		\hline
	\end{tabular}
\end{table*}

Second, the binary difference relation $w_1 - w_2 = b_{12}(m_2 - m_1)\,g_0/k_\mathrm{B}T$ is composition-independent. The crossover relation of \citet{Hunten1987} therefore holds within the framework at any heavy-gas abundance. The received claim that it requires trace heavies~\citep{Dauphas2014, Tian2018, Gronoff2020} has no basis in the algebra, and Chassefi\`ere had already applied it at an oxygen-to-hydrogen ratio of one half~\citep{Chassefiere1996a}. Third, the three-species system of \citet{Zahnle2023}. Their Eq.~(20) is derived in a regime where CO$_2$ is unaffected, so no CO$_2$ flux appears in it, and in their all-escaping regime (c$''$) they state the system can be solved without re-listing its members. Either way the case is not closed: Eq.~(20) is exact only while the CO$_2$ flux vanishes, and without it there are two equations for three unknowns. We checked this at a fully non-trace configuration, where the two sides agree to ten digits while CO$_2$ is retained and their ratio runs 0.951, 0.492, and $-1.117$ as the flux rises through entrainment. The correct difference relation adds the coupling $\phi_4 b_{12}(1/b_{14} - 1/b_{24})$, which the closure supplies for any escaping set. Their Eq.~(19) is the closure's multibackground activation threshold at $N = 3$, and their Eq.~(9) is the antecedent of the mole-fraction restatement, both recovered as identities (Table~\ref{tab:s_reductions}).

For a trace species $i$ riding on escaping set $\mathcal{A}$ with retained set $\mathcal{R}$, the closure reduces to

\begin{equation}
	\begin{aligned}
		w_i &= \frac{\sum_{j \in \mathcal{A}} X_j w_j / b_{ij} \;-\; \left(m_i g_0/k_\mathrm{B}T - 1/\bar{H}\right)}{\sum_{j \in \mathcal{A} \cup \mathcal{R}} X_j / b_{ij}},\\[2pt]
		&\mbox{entrained if } w_i > 0,
	\end{aligned}
	\label{eq:s_tracerider}
\end{equation}

\noindent generalizing Eqs.~(35), (36), and (44) of \citet{Zahnle1990} and Eq.~(5) of \citet{Odert2018} to arbitrarily many majors and backgrounds with all pairwise couplings, and the activation threshold of candidate $k$ when only species $\ell$ escapes is

\begin{equation}
	\Phi_\ell^*(k) = \frac{X_\ell\,(m_k - m_\ell)\, g_0/k_\mathrm{B}T}{(X_\ell + X_k)/b_{\ell k} + \sum_{k' \notin \{\ell, k\}} X_{k'}/b_{\ell k'}},
	\label{eq:s_threshold}
\end{equation}

\noindent reducing to the crossover condition of \citet{Hunten1987} for one background and to Eq.~(42) of \citet{Zahnle1990} for two retained heavies. The ternary reductions recover the isotope systems of \citet{Gu2023} and~\citep{Cherubim2024} term for term, including the helium effect (retained He steepens the common gradient and lowers the deuterium threshold, their zero crossings). The back-reaction term that~\citep{Gu2023} derive and discard is the $w_3$ portion of the exact $O(X_3)$ coupling, which the closure retains.

\subsection*{The diffusion-coefficient library and its error model}

The library (Data S1) covers all 91 pairs of the 14-species set \{H, He, C, N, O, Ne, Na, Mg, Si, S, Ar, Fe, Kr, Xe\}. Measured rows are the flow-tube compilations~\citep{Zahnle1986, Marrero1972} (assigned a 10\% uncertainty). Estimated rows follow the reduced-mass hard-sphere scaling with van der Waals diameters~\citep{Bondi1964, Batsanov2001, Alvarez2013} (a 30\% uncertainty), all sharing the single temperature exponent $T^{0.75}$ of the compilation of \citet{Zahnle2023}, whose printed CO$_2$--D entry carries a factor-of-ten misprint against its own scaling rule and is not propagated. No source prints atomic C, N, or S diameters, and those rows carry the widest uncertainty. The flagged spread of the printed Mg radius (36 to 55\% in $b$) is only recorded and not propagated, magnesium resting in any case on the neutral-species exception below. The one atomic-nitrogen system in print (N--N$_2$; \citealt{Marrero1972}) shares its recommended correlation identically with O--N$_2$ and O--O$_2$, so the measured record does not separate atomic N from O against a common partner beyond the mass factor. The row is still carried as a cross-check and the N/O pair prints with its width.

Because most estimated rows are scaled from the same few anchors, their ratios are far better determined than their values. The adopted error model is therefore correlated: each Monte Carlo draw perturbs the shared anchors and the per-row scatter separately (2000 draws per composition), and a pair verdict must hold under either choice of which measured rows anchor the scaling. The independent-per-row model is quoted once in the main text as a sensitivity and enters no figure or count. 

\subsection*{The independent transonic multifluid integrator and the validation protocol}

The validation integrator is a purpose-built one-dimensional, time-dependent, multifluid isothermal wind code relaxed to steady state, sharing the closure's inherited frame (one temperature, neutral gases, no chemistry). Nevertheless, it does not carry the two approximations under test: the subsonic setting of flux ratios and constant co-escaping composition. The simulation shares no code with the closure, one comparison script being the only place the two meet, and its own tests compare it against analytic and published results only, including the static-heavy configuration of \citet{Zahnle1990} (reproduced to better than 1.7\% by three-grid Richardson extrapolation at heavy-to-light mixing ratios of 0.3 to 3) and the two-species threshold identity of \citet{Zahnle1986}, whose predicted fractionation the integrator meets to within a percent in flux at the closure's own activation threshold.

\begin{figure*}[t]
	\centering
	\includegraphics[width=0.84\textwidth]{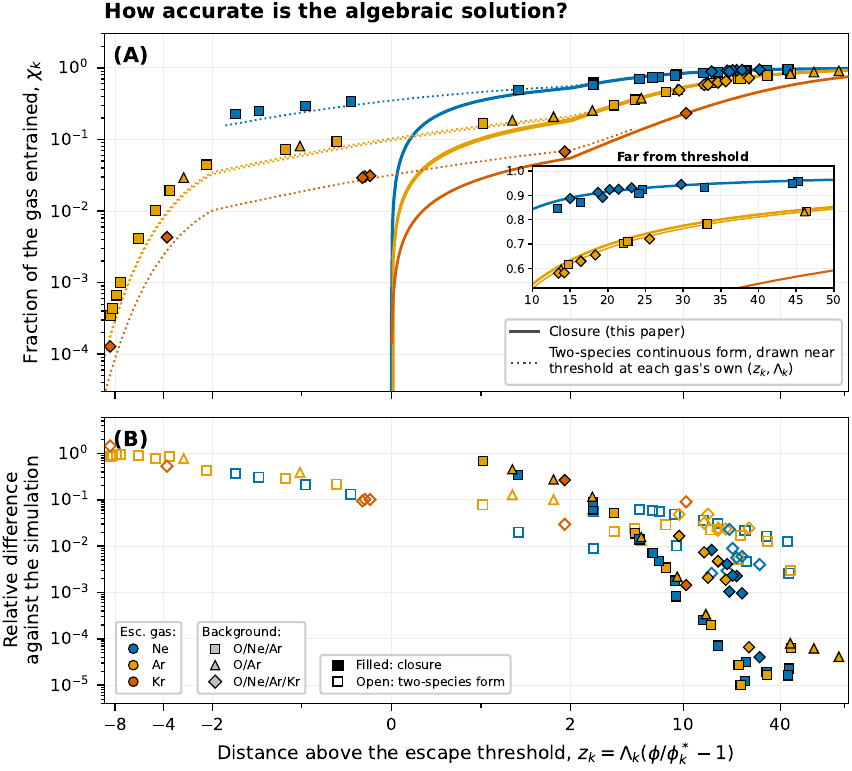}
	\caption{\textbf{The algebraic solution measured against an independent transonic multifluid simulation.}
		(\textbf{A})~Fraction of each gas entrained, $\chi_k$, against its distance from its own escape threshold, $z_k = \Lambda_k(\phi/\phi_k^* - 1)$: positive $z_k$ means the wind is that far above the gas' threshold, negative below it, in units of the retention slope $\Lambda_k$. Color gives the escaping gas (Ne blue, Ar orange, Kr vermillion) and marker the background mixture (squares, O/Ne/Ar; triangles, the two-gas O/Ar control; diamonds, the four-gas O/Ne/Ar/Kr control). Solid curves are the closed solution. Dotted curves are the published two-species continuous form~\citep{Zahnle1986}, drawn near threshold at each gas' own $(z_k, \Lambda_k)$, which the closure supplies and of which no four-gas version exists. The inset enlarges $10 < z_k < 50$ on linear axes.
		(\textbf{B})~Relative difference against the simulations, for the closure (filled symbols) and for the transplanted two-species form (open symbols). With every gas far from its own threshold ($z_k \gtrsim 15$) the closure agrees to a few parts in ten thousand, two to three orders of magnitude closer than the two-species form. Below threshold the measured tail floors at $\exp(-\Lambda_k)$, where the closure sets it to zero, and the transplanted form tracks it within factors of 1.1 to 2.5.}
	\label{fig:validation}
\end{figure*}

Every quoted accuracy is a Richardson extrapolation of a sequence of three grid resolutions (fitted convergence orders 1.9 to 3.1), repeated at neighboring configurations, with convergence behavior and grid resolutions recorded per run. From 67 accepted mixture runs, the two-species control runs, and 21 four-gas runs: with every gas far from its own threshold ($z_k \gtrsim 15$), the fractionation factors of simulation and closure agree to better than $3.5\times10^{-4}$, with mixed signs, over two to four species, mass ratios of 2 to 16, restricted Jeans parameters of 5 to 10, and heavy-gas mass loadings up to about one third. This is the verified domain of the main text (Fig.~\ref{fig:validation}). Which gas is the lightest does not enter: the dimensionless problem depends only on the mass ratios, $\lambda_0$, the drag parameters, and the mole fractions. Consequently, a run whose gases all carry sixteen times the mass of its twin's reproduces that twin to $10^{-14}$, and a wind with oxygen as its lightest gas returns the same residuals as the hydrogen-carried runs. Membership in the domain is a joint condition, and the four-gas O/Ne/Ar/Kr control (diamonds in Fig.~\ref{fig:validation}) shows it in both directions. When krypton is held just below its own threshold, the other three gases disagree with the closure by a few parts in a thousand even though each sits far above its own threshold ($z_k = 15$ to 21). The excess error enters through the drag terms coupling them to krypton, since tripling or thirding krypton's estimated coefficients moves the companions' residuals and flips one sign while leaving krypton's own escape nearly unchanged. When all four gases sit far above their thresholds, the residuals return to the $10^{-4}$ level ($-4.0$ and $-6.6\times10^{-5}$ for neon and argon). Below threshold, the converged runs measure the tails behind the factors of 1.1 to 2.5 quoted for Eq.~\eqref{eq:s_tail} (Fig.~\ref{fig:validation}A). The largest converged discrepancy in any accepted run is a 68\% understatement at $z = 1.02$. Representative measurements are in Table~\ref{tab:s_validation}. At heavy-gas mass loadings above about 0.6 the simulation reached no steady state from either of two opposed initial conditions, so nothing is claimed there in either direction.

\begin{table*}[t]
	\centering
	\caption{\textbf{Representative validation measurements against the independent transonic integrator.} Signed relative difference $d = (\chi_\mathrm{int} - \chi_\mathrm{clo})/\chi_\mathrm{clo}$ between the integrator and the closure, each a Richardson extrapolation of a grid-refinement triple (fitted orders 1.9 to 3.1). The O/Ne/Ar mixture carries neon at the stated mole fraction with argon in trace. The two-gas O/Ar rows are the binary control, and the four-gas rows are the O/Ne/Ar/Kr control with nothing trace. With every gas far from its own threshold, at $z_k \gtrsim 15$, every residual is below $3.5\times10^{-4}$.}
	\label{tab:s_validation}

	\small
	\begin{tabular}{l l r r}
		\\
		\hline
		Configuration & Gas & $z_k$ & $d$\\
		\hline
		O/Ne/Ar, $X_\mathrm{Ne} = 0.20$ & Ne & 1.42 & $+3.4\times10^{-1}$\\
		O/Ne/Ar, $X_\mathrm{Ne} = 0.20$ & Ne & 2.76 & $+9.1\times10^{-2}$\\
		O/Ne/Ar, $X_\mathrm{Ne} = 0.20$ & Ne & 5.35 & $+1.4\times10^{-2}$\\
		O/Ne/Ar, $X_\mathrm{Ne} = 0.20$ & Ne & 8.9 & $+1.8\times10^{-3}$\\
		O/Ne/Ar, $X_\mathrm{Ne} = 0.20$ & Ne & 13.3 & $+2.5\times10^{-4}$\\
		O/Ne/Ar, $X_\mathrm{Ne} = 0.20$ & Ne & 16.4 & $+7.2\times10^{-5}$\\
		O/Ne/Ar, $X_\mathrm{Ne} = 0.20$ & Ne & 24.4 & $-1.2\times10^{-5}$\\
		O/Ne/Ar, $X_\mathrm{Ne} = 0.20$ & Ne & 44.9 & $-1.6\times10^{-5}$\\
		O/Ne/Ar, $X_\mathrm{Ne} = 0.05$ & Ne & 8.9 & $+8.3\times10^{-4}$\\
		O/Ne/Ar, $X_\mathrm{Ne} = 0.05$ & Ne & 24.4 & $-3.2\times10^{-5}$\\
		Two-gas O/Ar control & Ar & 1.35 & $+4.6\times10^{-1}$\\
		Two-gas O/Ar control & Ar & 5.49 & $+1.6\times10^{-2}$\\
		Two-gas O/Ar control & Ar & 13.8 & $+3.4\times10^{-4}$\\
		Two-gas O/Ar control & Ar & 46.1 & $-8.0\times10^{-5}$\\
		Four-gas, Kr retained at $z_\mathrm{Kr} = -0.3$ & Ne, Ar & 15--21 & $-2.3$ to $-4.8\times10^{-3}$\\
		Four-gas, all entrained, $z_\mathrm{Kr} = +10.4$ & Ne & 29.7 & $-4.0\times10^{-5}$\\
		Four-gas, all entrained, $z_\mathrm{Kr} = +10.4$ & Ar & 25.6 & $-6.6\times10^{-5}$\\
		Four-gas, all entrained, $z_\mathrm{Kr} = +10.4$ & Kr & 10.4 & $+1.4\times10^{-3}$\\
		\hline
	\end{tabular}
\end{table*}

\subsection*{Archetypes, thresholds, and the sorting measurements}

Three archetype compositions carry the results (Table~\ref{tab:s_archetypes}): a primary atmosphere (He/H = 0.0820 by number, from the photospheric mass fractions of \citealt{Asplund2021}), a secondary atmosphere (atomized oxygen bulk with hydrogen pinned at mole fraction 0.01), and a rock vapor (the published mineral-vapor composition of \citealt{Ito2021} with the same trace-hydrogen carrier). Minor species sit at a $10^{-3}$ placeholder abundance, stated wherever a number depends on it. Activation thresholds are located by bisection on the closure. The ordering results are invariant where the algebra requires them to be: threshold ratios are independent of $\lambda_0$ and the base radius, and the library's single temperature exponent makes all thresholds shift together as $T^{-0.25}$. The sodium and magnesium thresholds rest on an explicit neutral-species exception, both ionizing readily at rock-vapor temperatures.

\begin{table*}[t]
	\centering
	\caption{\textbf{The three archetype atmospheres: thresholds, floor coefficients, and carriers.} Activation thresholds $\phi_k^*$ relative to each atmosphere's lowest threshold, the floor coefficient $\Lambda_k/\lambda_0$ (the sub-threshold floor is $\exp(-\Lambda_k)$, with $\lambda_0$ evaluated for each atmosphere's reference gas at the stated frame), and the flux-weighted mean mass $\bar{m}_\mathrm{w}$ of the wind just below each threshold. Primary atmosphere: He/H = 0.0820~\citep{Asplund2021}, reference gas H. Secondary atmosphere: atomized O bulk, H at 0.01, reference gas O. Rock vapor: mineral composition of \citet{Ito2021} with H at 0.01, reference gas O. Minor species at the $10^{-3}$ placeholder.}
	\label{tab:s_archetypes}

	\small
	\begin{tabular}{l rrr rrr rrr}
		\\
		\hline
		 & \multicolumn{3}{c}{Primary atmosphere} & \multicolumn{3}{c}{Secondary atmosphere} & \multicolumn{3}{c}{Rock vapor}\\
		Gas & $\phi_k^*$ & $\Lambda_k/\lambda_0$ & $\bar{m}_\mathrm{w}$ & $\phi_k^*$ & $\Lambda_k/\lambda_0$ & $\bar{m}_\mathrm{w}$ & $\phi_k^*$ & $\Lambda_k/\lambda_0$ & $\bar{m}_\mathrm{w}$\\
		\hline
		He & 1 & 2.97 & 1.0 & 1 & 0.187 & 1.0 & 1 & 0.187 & 1.0\\
		C & 2.58 & 10.1 & 1.1 & 4.36 & 0.649 & 1.2 & 4.28 & 0.652 & 1.1\\
		N & 3.40 & 12.1 & 1.2 & 5.47 & 0.620 & 1.2 & 5.30 & 0.653 & 1.2\\
		O & 3.40 & 13.8 & 1.2 & 6.87 & 0.604 & 1.3 & 6.53 & 0.655 & 1.3\\
		Ne & 4.56 & 17.4 & 1.2 & 698 & 0.252 & 14.3 & 302 & 0.255 & 12.2\\
		Na & 3.75 & 20.4 & 1.2 & 741 & 0.437 & 14.3 & 421 & 0.440 & 13.0\\
		Mg & 5.56 & 21.3 & 1.2 & $1.19\times10^3$ & 0.522 & 14.9 & 665 & 0.383 & 14.5\\
		Si & 5.15 & 25.1 & 1.2 & $1.35\times10^3$ & 0.756 & 15.0 & 900 & 0.618 & 15.4\\
		S & 7.22 & 28.2 & 1.3 & $2.09\times10^3$ & 1.01 & 15.3 & $1.49\times10^3$ & 0.790 & 17.0\\
		Ar & 6.07 & 31.6 & 1.2 & $2.09\times10^3$ & 1.25 & 15.3 & $1.77\times10^3$ & 0.995 & 17.4\\
		Fe & 9.02 & 50.2 & 1.3 & $3.38\times10^3$ & 2.49 & 15.5 & $2.89\times10^3$ & 2.29 & 18.3\\
		Kr & 18.1 & 77.8 & 1.4 & $7.05\times10^3$ & 4.23 & 15.7 & $6.02\times10^3$ & 4.04 & 19.2\\
		Xe & 25.9 & 121 & 1.4 & $1.04\times10^4$ & 7.19 & 15.8 & $9.13\times10^3$ & 7.00 & 19.5\\
		\hline
	\end{tabular}
\end{table*}

The gap measurements of the main text: the widest gap sits immediately above the lightest abundant gas in all five compositions with species heavier than their bulk, the xenon-dominated null case collapses the sequence's span from above $10^4$ to 252 with no wide gap, the carbon threshold is proportional to the carrier abundance to 1.0\% over four decades while the xenon threshold is constant to 2.4\%, and the hydrogen scan gives the quoted widths 368, 291, 102, and 14. A mass-loading estimate (the flux needed to accelerate the bulk at the gap's location) bounds every measured width from above, by factors of 1.33 to 8.27, and is quoted only as that bound. The shared thermosphere temperature is 5000~K ($\lambda_0(\mathrm{O}) = 24.1$ at Earth size), inside the 3500 to $10^4$~K span of published thermosphere models~\citep{Ito2021, Kite2020, Cherubim2026}, and the sub-threshold floor verdicts are exponentially sensitive to that choice while the ordering is not.

\subsection*{The retention plane and the planet sample}

Each threshold flux converts to a boundary in instellation against escape velocity through an energy-limited prescription, with the prescription factors varied within the following brackets: heating efficiency 0.10 to 0.30 for volatile atmospheres (or  $2.2\times10^{-4}$ to $3.8\times10^{-3}$ for rock vapor; \citealt{Ito2021}), XUV absorption radius 1.0 to 1.4 base radii, XUV-to-bolometric fraction $10^{-3.5}$ to $10^{-3.2}$, and bulk densities 3.0 to 8.0 g\,cm$^{-3}$. The combined bracket is the factor of about 12 of the main text, common to all boundaries. The family reduces to the energy-limited shoreline prediction of \citet{Zahnle2017}, their Eqs.~(33) and (34), XUV instellation proportional to $v_\mathrm{esc}^3\sqrt{\rho}$, and not to their empirical $v_\mathrm{esc}^4$ line, which their figure caption describes as drawn by eye and their abstract as differing distinctly from the prediction. The species-resolved family shifts that prediction gas by gas but leaves its slope unchanged, so it does not bear on the tension between the predicted scaling and the steeper empirical line. The reduction is exact, and the two sides of it are independent, which is what keeps it from being circular. On our side every threshold is proportional to $g_0$, which enters Eqs.~\eqref{eq:s_system} and~\eqref{eq:s_threshold} only through the driving term $m_j g_0/k_\mathrm{B}T$. The energy-limited prescription gives $\phi \propto F_\mathrm{XUV}/v_\mathrm{esc}^2$, and a fixed bulk density gives $r_0 \propto v_\mathrm{esc}/\sqrt{\rho}$, so that $F^*_{\mathrm{XUV},k} \propto v_\mathrm{esc}^2 g_0 \propto v_\mathrm{esc}^4/r_0 \propto v_\mathrm{esc}^3\sqrt{\rho}$. Zahnle and Catling reach the same pair from a different requirement: their Eq.~(32) gives the fraction of a planet's mass removed over an integrated XUV history, and holding that fraction fixed holds fixed a removed column proportional to $\rho r_0$, which is the combination our entrainment threshold carries for an unrelated, diffusive reason. The prescription is the one ingredient the two routes share, supplying the factor $v_\mathrm{esc}^2$ in both. No shoreline datum enters the boundary family, and nothing in it is fitted. The closure's binary activation threshold reproduces the published crossover flux of \citet{Kite2020} to $10^{-12}$ relative through an independent conversion.

The seventeen-planet sample (Table~\ref{tab:s_planets}) is the nine-planet thermal-emission sample of \citet{Coy2025}, taken from their Tables 1 and 2, and the nine Rocky Worlds DDT targets~\citep{Redfield2024}\footnote{\url{https://rockyworlds.stsci.edu/}}, of which LTT~1445~A~b is already in the emission sample, the other eight each placed by a named refereed characterization: mass, radius, and instellation with 1$\sigma$ widths from that reference's parameter set. The adopted parameters agree with the program page's to within 15\%, except where a newer characterization supersedes the page: GJ~3929~b~\citep{Beard2022}, TOI-198~b~\citep{ZapateroOsorio2026}, and TOI-771~b~\citep{Lacedelli2025}. For LTT~1445~A~c we adopt~\citep{Pass2023}, whose space-based photometry resolved the transit's grazing geometry. The earlier radius of \citet{Winters2022}, still the default in catalog compilations, is a lower limit set by that geometry. The retention index of planet $p$ is $\mathcal{R}_p = F_{\mathrm{XUV},p}\, /\, F^*_{\mathrm{XUV,S}}(v_{\mathrm{esc},p})$: the XUV flux the planet receives over the flux at which a wind at its escape velocity starts stripping the sulfur-bearing gases, so that $\mathcal{R}_p > 1$ predicts stripping and $\mathcal{R}_p < 1$ retention. Every prescription factor multiplies all seventeen indices together, so the ranking is prescription-free while the absolute crossing carries the factor-of-12 bracket. TOI-198~b sits within its instellation width of the boundary ($S = 7.2 \pm 2.4$ Earth insolations against an index of 0.88) and is counted on neither side.

\begin{table*}[t]
	\centering
	\caption{\textbf{The seventeen-planet sample with its retention indices.} Masses and radii in Earth units, instellation $S$ in Earth insolations, escape velocity in km\,s$^{-1}$, and the retention index (the XUV-flux margin against the sulfur boundary of the steam-residue archetype at the stated 5000~K frame, prescription factors multiply all indices together). All seventeen rows carry 1$\sigma$ widths: the nine emission-sample planets take theirs from Tables 1 and 2 of \citet{Coy2025} (the LHS 3844 b mass is unmeasured, and the $\pm 1.0$ is the adopted value's stated width), and the eight Rocky Worlds DDT rows not shared with the emission sample take each row from the named refereed characterization. TOI-198~b sits within its instellation width of the boundary and is counted on neither side.}
	\label{tab:s_planets}

	\small
	\begin{tabular}{l rrr r r l}
		\\
		\hline
		Planet & $M_\mathrm{p}$ & $R_\mathrm{p}$ & $S$ & $v_\mathrm{esc}$ & Index & Source\\
		\hline
		GJ 367 b & $0.63 \pm 0.05$ & $0.70 \pm 0.02$ & $578 \pm 54$ & $10.61 \pm 0.45$ & 244 & \citet{Coy2025}\\
		GJ 1252 b & $1.32 \pm 0.28$ & $1.18 \pm 0.08$ & $234 \pm 60$ & $11.83 \pm 1.32$ & 108 & \citet{Coy2025}\\
		TOI-1685 b & $3.03 \pm 0.33$ & $1.38 \pm 0.04$ & $235 \pm 24$ & $16.57 \pm 0.93$ & 33 & \citet{Coy2025}\\
		LHS 3844 b & $2.2 \pm 1.0$ & $1.30 \pm 0.02$ & $69.9 \pm 6.9$ & $14.55 \pm 3.31$ & 15 & \citet{Coy2025}\\
		GJ 486 b & $2.77 \pm 0.07$ & $1.29 \pm 0.02$ & $39.2 \pm 1.6$ & $16.39 \pm 0.24$ & 5.4 & \citet{Coy2025}\\
		GJ 1132 b & $1.84 \pm 0.19$ & $1.19 \pm 0.04$ & $19.4 \pm 1.3$ & $13.91 \pm 0.76$ & 4.7 & \citet{Coy2025}\\
		TRAPPIST-1 b & $1.37 \pm 0.07$ & $1.12 \pm 0.01$ & $4.16 \pm 0.15$ & $12.37 \pm 0.32$ & 1.5 & \citet{Coy2025}\\
		TRAPPIST-1 c & $1.31 \pm 0.06$ & $1.10 \pm 0.01$ & $2.21 \pm 0.09$ & $12.21 \pm 0.28$ & 0.84 & \citet{Coy2025}\\
		LTT 1445 A b & $2.70 \pm 0.20$ & $1.30 \pm 0.10$ & $5.4 \pm 1.1$ & $16.12 \pm 0.86$ & 0.79 & \citet{Coy2025}\\
		\hline
		TOI-406 c & $2.08 \pm 0.23$ & $1.32 \pm 0.12$ & $19.0 \pm 3.0$ & $14.04 \pm 1.01$ & 4.9 & \citet{Lacedelli2024}\\
		LTT 1445 A c & $1.37 \pm 0.19$ & $1.07 \pm 0.10$ & $11.7 \pm 2.7$ & $12.66 \pm 1.06$ & 3.7 & \citet{Pass2023}\\
		GJ 3929 b & $1.75 \pm 0.44$ & $1.09 \pm 0.04$ & $17.3 \pm 0.8$ & $14.17 \pm 1.80$ & 3.6 & \citet{Beard2022}\\
		HD 260655 c & $3.09 \pm 0.48$ & $1.53 \pm 0.05$ & $16.1 \pm 0.3$ & $15.88 \pm 1.26$ & 3.0 & \citet{Luque2022}\\
		TOI-771 b & $2.47 \pm 0.32$ & $1.36 \pm 0.10$ & $14.0 \pm 4.0$ & $15.07 \pm 1.12$ & 2.8 & \citet{Lacedelli2025}\\
		TOI-244 b & $2.68 \pm 0.30$ & $1.52 \pm 0.12$ & $7.3 \pm 0.4$ & $14.85 \pm 1.02$ & 1.7 & \citet{CastroGonzalez2023}\\
		TOI-198 b & $3.17 \pm 0.64$ & $1.36 \pm 0.13$ & $7.2 \pm 2.4$ & $17.08 \pm 1.91$ & 0.88 & \citet{ZapateroOsorio2026}\\
		LHS 1140 b & $5.60 \pm 0.19$ & $1.73 \pm 0.03$ & $0.43 \pm 0.03$ & $20.12 \pm 0.37$ & 0.035 & \citet{Cadieux2024}\\
		\hline
	\end{tabular}
\end{table*}

A condensation screen accompanies the plane: on this sample no element of the sequence is cold-trapped at the relevant levels, and a static trap for a condensible held at mixing ratio $10^{-6}$ over 190 to 250~K would require 0.33 to 762~bar of diluent gas\footnote{NIST Chemistry WebBook, NIST Standard Reference Database Number 69, \url{https://webbook.nist.gov/}}, so retention statements on this sample are entrainment statements. The fluid-regime constraint is stated under both applications in Table~\ref{tab:s_regime}. The published transition out of the fluid regime sits at $\lambda_0 \approx 3$~\citep{Volkov2011, Johnson2013a, Johnson2013b}, loosened here to 6 and 12 to cover the conversion from the top of their heated region to the closure's base level. Applied to the oxygen bulk, it admits the sharply dropping thresholds only at its loose end: the minimum entrained masses are 53.3, 34.7, and 25.3 amu in an oxygen wind at ceilings of 3, 6, and 12. Resolved by carrier, the gas escaping just below the C, N, and O thresholds is hydrogen, at flux-weighted mean mass 1.15 to 1.30 and $\lambda_0(\mathrm{H}) = 1.4$ to 3.3 at the stated frame. Every oxygen-carried boundary stands on the same extrapolation as the energy-limited shoreline literature, the published transition calculations having no heating above the sonic point by construction~\citep{Volkov2011}. Neither application is preferred here.

\begin{table*}[t]
	\centering
	\caption{\textbf{The fluid-regime constraint under its two applications.} The published single-species transition out of the fluid regime~\citep{Volkov2011, Johnson2013a, Johnson2013b} sets a ceiling on the restricted Jeans parameter $\lambda_0$. That transition is printed near $\lambda_0 = 3$. The ceilings 6 and 12 are loosened from it here to cover the conversion to the closure's base level. On a multispecies wind the ceiling can apply to the static bulk or to the gas actually escaping, and the two applications admit complementary parts of the threshold sequence. Neither is preferred here.}
	\label{tab:s_regime}

	\small
	\begin{tabular}{p{0.28\textwidth}p{0.32\textwidth}p{0.32\textwidth}}
		\\
		\hline
		Applied to & Quantity constrained & Consequence on this sample\\
		\hline
		Bulk: ceiling on the oxygen background & $\lambda_0(\mathrm{O})$ at ceilings 3, 6, 12 & minimum sharply dropping threshold masses 53.3, 34.7, 25.3 amu; only the S/Ar boundary and above survive, at the loose end of the ceiling\\
		Carrier-resolved: ceiling on the escaping gas & flux-weighted mean mass of the wind below each threshold & C, N, O boundaries are hydrogen-carried ($\bar{m}_\mathrm{w} = 1.15$ to 1.30, $\lambda_0(\mathrm{H}) = 1.4$ to 3.3 at 5000~K), inside the published fluid range on the whole sample; every oxygen-carried boundary stands on the shoreline literature's shared extrapolation\\
		\hline
	\end{tabular}
\end{table*}


\section*{Supplementary Text}

\subsection*{Validity domain and inherited assumptions}

The closure inherits the assumptions of the algebraic formalisms it generalizes. The wind is steady, spherically symmetric, isothermal, neutral, and chemistry-free. The energy budget lives entirely in the prescribed flux. The relative fluxes are set in the subsonic region and carried outward, a carry-over whose binary record is the transonic comparison of \citet{Zahnle1986} and whose breakdown (buoyant lofting above a retained heavy background at base mixing ratios of order unity) is documented at figure level in~\citep{Zahnle1990}. Our validation quantifies the carry-over in the verified domain and inherits the breakdown thresholds outside it. Hydrodynamic validity of the whole picture requires the retained-heavy loading to satisfy roughly $f \lesssim 3$~\citep{Zahnle1990}, which the implementation cleanly reports as a diagnostic. The kinetic transition out of the fluid regime is mapped for single-species winds by \citet{Volkov2011, Johnson2013a, Johnson2013b, Evans2025}, and the unheated bound $\lambda_0 \lesssim 3$ of \citet{Volkov2011} excludes every configuration in this paper and in the energy-limited literature alike, a shared extrapolation the main text declares. All quantities are evaluated at the XUV base with atomized composition, an upper-limit convention on photolysis with a known bias direction~\citep{Wordsworth2018}. The base is one level for all species, tide-corrected gravity passes through unchanged, and the handover to Jeans escape at low flux belongs to the escape-rate prescription~\citep{Chassefiere1996b, Yelle2024}. The one omission that pushes against the upper-limit framing is the neutral-only $b_{ij}$: coupling to ions is stronger, so drag-driven entrainment is underestimated, and the ionization corrections of \citet{Gu2023} are the documented refinement. The escaping composition is dynamically stable within a layer-relaxation model: strictly proven for any binary at any composition, and supported for $N \ge 3$ by 2000-draw Jacobian ensembles with no unstable case found.

\subsection*{Relation to neighboring formalisms}

The Stefan--Maxwell solvers of planetary aeronomy~\citep{GarciaMunoz2007, Ern1994} and the drag-matrix integrators of multispecies dynamics~\citep{BenitezLlambay2019} solve the prescribed-composition diffusion problem the closure does not, and vice versa: there the composition profile is the unknown and the fluxes follow, here the bulk flux is prescribed and the partition and escaping set are the unknowns. The nearest analytic treatment~\citep{Erkaev2026} is an asymptotic expansion for trace minors in a designated hydrogen wind, solved as a $2\times2$ linear system by Cramer's rule. Escape fluxes for multispecies winds are otherwise obtained by full numerical solution instead of self-consistent algebra~\citep{Evans2025}. The same force balance under a different drag, Coulomb collisions in place of neutral collisions, fractionates the solar wind~\citep{Geiss1982, Bodmer2000}, and prescribed-carrier-flux fractionation has an independent life in geochemistry~\citep{Severinghaus1996}.

\subsection*{The early-Mars case and its conditions}

The early-Mars case is the first of the main text's confrontations with a measured record. The escape attribution of the meteorite argon is stated by \citet{Willett2022}, with the trapped $^{38}$Ar/$^{36}$Ar $= 0.41 \pm 0.05$ as an anchor against a solar $0.1818$. That paper's corrigendum~\citep{Willett2023} replaces a supplementary table without amending any main-text value. What the closure adds to that record is an ordered sequence. Such ordering is the part that survives every variation we tested: carbon's entrainment threshold sits below krypton's at $P \ge 0.994$ across four compositions spanning hydrogen-rich to oxygen-loaded, both carbon speciations (CO$_2$ and CO), and both coefficient-uncertainty models, at 2000 draws per case. The stronger version of the same statement, that any wind entraining argon is already carrying carbon, holds only in the heavy-loaded compositions. Argon and carbon dioxide, for their part, cannot be ordered against each other at all: their thresholds are undecided by the coefficient library in every case tested, the same pattern as the S/Ar and N/O pairs above.

The quantitative loss bound is a statement over what was explored rather than a proof over all possible monotone histories. Two families enter it, constant flux and threshold-tracking, over 28 flux factors, four compositions, and both speciations, of which 332 of 448 tracks converged, and none of the remainder is near qualifying. Within that scan no history reproduces the argon record while sparing carbon, and none in the CO speciation reaches the measured enrichment at all. Because the exploration contains no replenishment, each loss statement bounds the escape phase itself. Krypton enters on a different footing. Its retention is a boundary statement rather than a marginal one at every stated frame ($\Lambda_\mathrm{Kr} = 51$ to $206$ at Mars radius over 500 to 2000~K, with floors below $10^{-22}$), but the krypton records themselves are modern or young-trapping~\citep{Swindle1986, Conrad2016}, and the familiar description of Martian krypton as unfractionated traces to a comparison against average carbonaceous chondrite rather than against solar, the interior baseline the Chassigny krypton measurements established \citep{Peron2022, Peron2025}. Those records are therefore consistent with the prediction without constraining the 4.4~Ga epoch from both sides. The case also inherits its assumptions, neutral gases throughout, which is why the xenon record of the same planet, an ion-driven story, lies outside it. The Archean-Earth counterpart of that boundary is quantified next.

\subsection*{The Archean-Earth xenon scan and its conditions}

The xenon statement of the main text has three layers, with bars fixed before any run on the template of the Mars exploration above. The ordering layer takes the conservative isotope pair, the heaviest krypton isotope ($^{86}$Kr) against the lightest xenon isotope ($^{130}$Xe), so that the printed probability covers every pair of the two elements. On the 2000-draw ensembles of the three archetypes, $P(\Phi^*_\mathrm{Kr} < \Phi^*_\mathrm{Xe}) = 0.996$ (secondary atmosphere), 1.000 (rock vapor), and 0.935 to 0.940 (primary atmosphere, undecided at the bar of 0.95). Five further compositions were built for this exploration, four spanning the published Archean ranges for N$_2$, CO$_2$, and total hydrogen~\citep{Catling2020} and one hydrogen-dominated end member carried as the most favorable neutral case. On those, $P \ge 0.9995$ in the four Archean compositions and 0.9875 in the end member under the correlated error model, with the independent-per-row model at 0.84 to 0.93 as the named sensitivity. The floor layer evaluates $\Lambda_\mathrm{Kr}$ and $\Lambda_\mathrm{Xe}$ at various conditions (500 to 2000~K, XUV level at 1.0 and 1.5 Earth radii, quoted on both the oxygen and the N$_2$ reference): the loosest frame gives $\Lambda_\mathrm{Kr} = 140$, so the krypton floor never exceeds $10^{-60}$ and the xenon floor ($\Lambda_\mathrm{Xe} \ge 259$) is smaller still, closing the neutral sub-threshold channel at any Archean epoch. The history layer scans two monotone flux families (constant, and threshold-tracking on $^{130}$Xe) at 28 flux factors per composition with reservoir bookkeeping only and step-halving convergence on every track. A converged history qualifies if its residual $^{136}$Xe/$^{130}$Xe enrichment reaches the measured band's 1$\sigma$ lower edge and its $^{130}$Xe loss reaches one half. The anchors are measured: the post-3.3-Ga increment of $12.9 \pm 1.2$ permil per amu (enrichment 0.080, 1$\sigma$ band 0.072 to 0.088, over the pair's 6 amu span) and the krypton record identical to modern at 3.3 and 2.45 Ga, which excludes the 16.2 and 8.3 permil per amu a mass-scaled extension of the xenon process would give~\citep{Avice2018}. The full air-to-solar slope of $36.2 \pm 1.6$ permil per amu (enrichment 0.238)~\citep{Dauphas2003} is carried as a second target.

Of 280 tracks, 277 converged, and the three that did not are far from qualifying in any case. Counting from the band's lower edge (enrichment 0.072), 80 histories qualify. Over all of them the krypton loss is at least 0.774, the ratio of krypton to xenon fractional loss is at least 1.07, the krypton isotopic enrichment is at least 13.6 permil per amu (against the 16.2 the samples exclude, and far from a record identical to modern), and the N$_2$ plus CO$_2$ loss is at least 0.946, against Archean N$_2$ bounds near or below the modern value~\citep{Catling2020, Avice2018}. At the central anchor (0.080, grid target) the corresponding minima over 69 histories are 0.774, 1.09, and 15.4 permil per amu. At the full air-to-solar target, and at 0.20 and 0.40, no scanned history qualifies at all. The violation sentence of the main text therefore rests on the elemental channel, which passes everywhere, with the isotopic minimum quoted as measured. The scan contains no replenishment from interior processes, so each loss statement bounds the escape phase itself. Additionally, the wind is modeled as neutral so the ion-coupled resolution~\citep{Zahnle2019, Catling2020} lies outside it. The xenon chronometer for the onset of volatile recycling into the mantle \citep{Parai2018} presumes the same prolonged ion-driven history, so the closed neutral channel supports its foundation. One circularity is avoided: the published Archean CH$_4$ floor above 5000 ppmv at 3.5 Ga rests on the xenon ion-drag requirement itself~\citep{Catling2020} and is not used here as an independent hydrogen constraint. 

\subsection*{The Venus argon scan and its conditions}

The Venus statement is explored the way the Mars and Archean-Earth ones are: four water-derived compositions spanning the published scenario~\citep{Zahnle2023} (stoichiometric steam over CO$_2$ at two CO$_2$ levels, an oxygen-enriched residue, and a late hydrogen-depleted stage), two monotone flux families at 28 factors each, and every history stopped when 95\% of its hydrogen and oxygen is gone, the epoch at which a water-powered wind loses its source. The measured box is the $^{38}$Ar/$^{36}$Ar of $0.183 \pm 0.003$~\citep{Avice2022} against solar~\citep{Willett2022}, an enrichment of $0.007 \pm 0.017$, together with an argon loss of at most one half, an assumption standing in for the unknown initial inventory. The looser published uncertainty of $\pm 0.02$ on the ratio admits any primordial end member and is carried only as a sensitivity (141 histories in that box against 97). In-situ sampling of the present upper atmosphere could tighten that box \citep{Borner2026}. Of 224 tracks, 208 converged and stripped the water, and 97 land in the box, at constant-flux factors up to 0.58 of argon's initial threshold in the CO$_2$-poorest composition and up to 0.11 to 0.15 in the others. Every one of the 97 removes no CO$_2$ and at most 3\% of the argon, a consistency with Venus keeping its CO$_2$ rather than a prediction of it. Every converged history at higher flux misses the box, 111 by over-fractionating the argon and 56 by removing more than half of it, and the sixteen tracks that failed their step-halving checks sit at the box's low-flux boundary or far outside it, blurring the boundary's location without touching either claim. A few additional considerations need to be accounted for. On the oxygen reference, the sub-threshold tail lies at $\Lambda_\mathrm{Ar} = 52$ to 309, so argon below its threshold is spared absolutely while the water-derived bulk lasts. On the CO$_2$ reference, $\Lambda_\mathrm{Ar}$ is negative, argon being lighter than CO$_2$, so a wind that keeps blowing after the water is gone strips the argon of any composition: in a first run of the exploration without the water-exhaustion stop, preserved with the results, no history landed in the box at all. 

\subsection*{The wrong-carrier factor}

On the published configuration of the helium-dominated candidate atmosphere of LHS~1140~b~\citep{Cherubim2026}, the binary crossover formula requires a designated carrier, and the computed oxygen escape rate depends on that designation by a factor of 2118 (designating helium gives $1.069\times10^9$ g\,s$^{-1}$, designating hydrogen $5.045\times10^5$ g\,s$^{-1}$). The closure needs no designation and returns $1.0695\times10^9$ g\,s$^{-1}$, confirming that the published choice was the correct one. This is the measurement behind the main text's factor of about 2000. It is not an error in the published number, but a decision that had to be made outside the calculation and gotten right by hand, which the simultaneous solution removes. 

That configuration is itself an inference from an escape diagnostic, excess transit absorption in the metastable helium triplet at 10830~\AA~\citep{Oklopcic2018}, and the observational record is still accumulating. Indeed, the ground-based signal was present in one transit and absent in a second~\citep{Cherubim2026}, and three independent analyses of four archival JWST NIRISS transits taken between 2023 and 2026, none of them contemporaneous with the ground-based detection, report no helium absorption and upper limits below the reported amplitude, while leaving episodic escape open~\citep{Bennett2026, Gressier2026, Radica2026}. We use the configuration as published and take no position on how that record resolves, because nothing here rests on it. The factor of 2118 is a property of the binary formula evaluated at a stated composition, and any helium-rich, hydrogen-poor mixture returns one of the same size. 

The retention plane is likewise indifferent: it places LHS~1140~b about seventy times above helium's entrainment boundary, a margin the prescription bracket of about 12 does not close, so helium is among the gases the plane does not expect the planet to keep, whether or not an outflow is carrying it at the present epoch.

\subsection*{Sulfur as the heaviest abundant gas of a secondary atmosphere}

The sample straddles the sulfur boundary, and sulfur is the heaviest element such an atmosphere holds in abundance for the purpose of the comparison. The rock-forming elements interleaved nearby in the sequence (Na, Mg, Si, and Fe) are excluded twice over: sodium and magnesium ionize readily at the temperatures where a rock vapor exists, while the closure is neutral, and the mineral species condense out of a steam-residue atmosphere at the sampled instellations. Sulfur-bearing gases survive both cuts and are observable in emission, and SO$_2$ has already been invoked to explain JWST spectra of WASP-39~b~\citep{Tsai2023} and L~98-59~d~\citep{Nicholls2026}. Above the sulfur boundary every major gas of the archetype is lost.


\runinlabel{Caption for Data S1.}
\textbf{The binary diffusion coefficient library.} Machine-readable table (CSV) of the 91 gas pairs of the 14-species set: pair, adopted $b_{ij}(T)$ coefficient and temperature exponent, provenance (measured row, scaled row, or fallback), source, uncertainty class, and per-row notes. The library and the code that regenerates it are part of the Zenodo deposit (\url{https://doi.org/10.5281/zenodo.22181501}).

\end{document}